\documentclass[useAMS,usenatbib,usegraphicx]{mn2e}
\usepackage{times}

\newcommand{\MCFOST}[0]{{\sc Mcfost}}
\newcommand{\ProDiMo}[0]{{\sc ProDiMo}}
\newcommand{\Tg}[0]{{T_{\rm g}}}
\newcommand{\Td}[0]{{T_{\rm d}}}
\title[Continuum and line modelling of discs around young stars
       \ I.\ 300\,000 disc models for Herschel/GASPS]
      {Continuum and line modelling of discs around young stars\\
       I.\ \ 300\,000 disc models for Herschel/GASPS}
\author[P.~Woitke et al.]
       {P.~Woitke$^{1,2}$, 
        C.~Pinte$^3$,
        I.~Tilling$^4$,
        F.~M{\'e}nard$^5$,
        I.~Kamp$^6$,
        W.-F.~Thi$^{4}$,
        \newauthor
        G.~Duch{\^e}ne$^{5,7}$,
    and J.-C.~Augereau$^5$\\
{\ }\\
$^1$ UK Astronomy Technology Centre, Royal Observatory, Edinburgh,
     Blackford Hill, Edinburgh EH9 3HJ, UK, {\tt email:\,ptw@roe.ac.uk}\\
$^2$ School of Physics \& Astronomy, University of St.~Andrews,
     North Haugh, St.~Andrews KY16 9SS, UK\\
$^3$ School of Physics, University of Exeter, Stocker Road, Exeter
     EX4 4QL,UK\\
$^4$ Institute for Astronomy, University of Edinburgh, Royal Observatory, 
     Blackford Hill, Edinburgh, EH9 3HJ, UK\\
$^5$ Laboratoire d'Astrophysique de Grenoble, CNRS/Universit{\'e} Joseph 
     Fourier (UMR5571) BP 53, F-38041 Grenoble cedex 9, France\\
$^6$ Kapteyn Astronomical Institute, Postbus 800, 9700 AV Groningen, 
     The Netherlands\\
$^7$ {Astronomy Department, University of California, Berkeley, 
     CA 94720-3411, USA}
}
  
\begin{document}

\date{Received 13.\ Feb.\ 2010;\ \  Accepted 11.\ Mar.\ 2010}

\pagerange{\pageref{firstpage}--\pageref{lastpage}} \pubyear{2010}

\maketitle

\label{firstpage}
\begin{abstract}
  We have combined the thermo-chemical disc code \ProDiMo\ with the
  Monte Carlo radiative transfer code \MCFOST\ to calculate a grid
  of $\sim$300\,000 circumstellar disc models, systematically varying
  11 stellar, disc and dust parameters including the total disc mass,
  several disc shape parameters and the dust-to-gas ratio. For each
  model, dust continuum and line radiative transfer calculations are
  carried out for {29} far IR, sub-mm and mm lines of [OI], [CII],
  $^{12}$CO and o/p-H$_2$O under 5 inclinations. {The grid allows
  to study the influence of the input parameters on the observables,
  to make statistical predictions for different types of circumstellar
  discs, and to find systematic trends and correlations between the
  parameters, the continuum fluxes, and the line fluxes. The model
  grid, comprising the calculated disc temperature and chemical
  structures, the computed SEDs, line fluxes and profiles, will be
  used in particular for the data interpretation of the {\sc Herschel}
  open time key programme GASPS. The calculated line fluxes show a
  strong dependence on the assumed UV excess of the central star, and
  on the disc flaring. The fraction of models predicting [OI] and
  [CII] fine-structure lines fluxes above {\sc Herschel/Pacs} and {\sc
  Spica/Safari} detection limits are calculated as function of disc
  mass. The possibility of deriving the disc gas mass from line
  observations is discussed.}
\end{abstract}

\begin{keywords}
          astrochemistry; 
          stars: formation; 
          circumstellar matter; 
          radiative transfer; 
          methods: numerical
\end{keywords}

\section{Introduction}

The {structure, composition and evolution of protoplanetary discs
are important corner-stones to unravel the mystery of life, as they
set the initial conditions for planet formation}. Spectral energy
distributions (SEDs), {although their analysis is known to be
degenerate}, probe the amount, temperature and overall geometry of the
dust in the discs, such as disc flaring \citep{Meeus2001},
puffed-up inner rims \citep{Dullemond2001, Acke2009}, and indications
of an average grain growth in discs as young as a few Myr
\citep{D'Alessio2001}.

Most works in the past decade have focused on the analysis of SEDs of
individual objects \citep[e.g.][]{D'Alessio2006} or to study
systematic trends in infrared colours of various types of discs,
ranging from embedded young stellar objects to exposed T\,Tauri stars
\citep{Robitaille2006}. Some ambiguities inherent in SED analysis can
be resolved by images in scattered light \citep{Stapelfeldt1998}, in
mid-infrared thermal emission \citep{McCabe2003} or in the mm-regime
\citep{Andrews2007}. The Spitzer observatory has enabled detailed
studies on dust mineralogy, constraining dust properties 
in the upper layers of the inner disk regions by using solid-state
features \citep{Furlan2006, Olofsson2009}.  
Multi-technique, panchromatic approaches, combining the aforementioned
observations, are now becoming possible but remain limited to a few
objects with complete data sets \citep[e.g.][]{Wolf2003, Pinte2008,
Duchene2009}.


\begin{figure*}
  \centering
  {\ }\\*[-2ex]
  \hspace*{-1mm}\includegraphics[width=16cm]{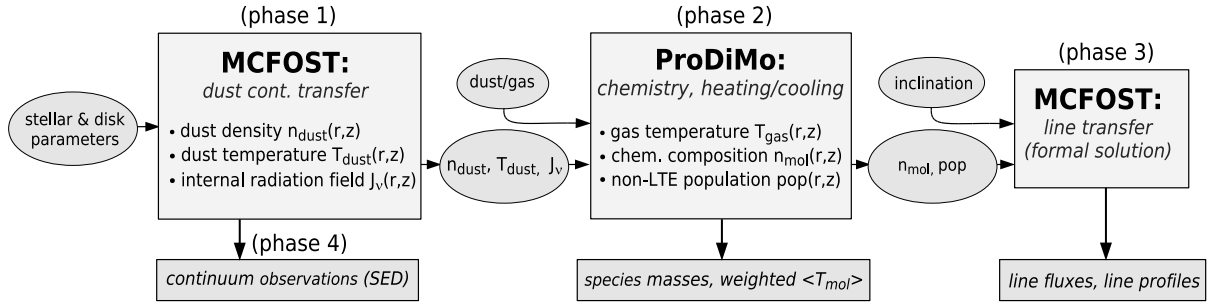}\\[-1mm]
  \caption{Unified {\sc Mcfost / ProDiMo} model pipeline to calculate one
    radiation thermo-chemical disc model with SED and line flux prediction.}
  \label{fig:ModelPipeline}
  \vspace*{-1mm}
\end{figure*}

However, all these observational findings are related to the dust
component. {Initially}, about 99\% of the disc mass can be assumed to
be present in form of gas, and the progress toward a better
understanding of the gas component, such as chemical composition, gas
temperature structure and vertical disc extension, is hampered by a
{current} lack of observational data. With several {key programs of
the {\sc Herschel Space Observatory}, such as GASPS and WISH, a new
body of gas emission line observations will be provided very soon, and
there is a clear need to include the gas in such systematic studies.}
Recent work has focused on the prediction of far IR line emissions
from individual discs
\citep{Meijerink2008,Ercolano2008,Woitke2009b,Cernicharo2009}, or
rather small parameter studies \citep{Kamp2009}.
\citet{Goicoechea2009} have recently discussed the detection rates of
the $\rm[OI]\,63\,\mu$m, $\rm[SI]\,56\,\mu$m and $\rm
[SiII]\,34\,\mu$m fine-structure lines with {\sc Herschel} and the
proposed {\sc Spica/Safari} mission on the basis of {\sl one} low mass
disc model ($M_{\rm gas}\!=\!10^{-5}\rm\,M_\odot$) by
\citet{Gorti2004}. {Their conclusions, however, depend on the choices
of the other model parameters, and it is difficult to put them on a
firm statistical basis.}

To address these issues, {we have combined our state-of-the-art
computer codes to calculate the dust and line radiative transfer with
the \MCFOST-code \citep{Pinte2006}, and the gas thermal
balance and chemistry with the \ProDiMo-code
\citep{Woitke2009a,Kamp2009}}.  We have computed a large grid of
300\,000 disc models to simultaneously predict SEDs and gas
emission lines from parametrised disc density distributions. The grid
name DENT stands for \underline{D}isc \underline{E}volution with
\underline{N}eat \underline{T}heory. The following sections describe
the model pipeline (Sect.~\ref{DENTgrid}) and the first results on
fine structure emission line fluxes and detectability
(Sect.~\ref{results}). Section \ref{summary} summarises the
preliminary findings of the DENT grid and provides an outlook to
future studies.

\section{The DENT grid}
\label{DENTgrid}

The DENT grid is a tool to investigate the influence of stellar, disc
and dust properties (Table~\ref{tab:Parameters}) on continuum and line
observations (Table~\ref{tab:Results}), and to study in how far these
dependencies can be inverted. The grid is designed to coarsely sample
the parameter space associated with young, intermediate to low-mass
stars at ages (1-30\,Myr) having gas-rich and gas-poor discs.

\begin{table}
\vspace*{-1mm}
\caption{Parameters of the DENT grid and values assumed. $R_{\rm
  subli}$ stands for the dust sublimation radius
  (where $\Td\!=\!1500\,$K). The choice of inclination angles resembles
  a randomly oriented sample.}
\label{tab:Parameters}
\vspace*{-1.5mm}
\hspace*{-2.5mm}\resizebox{8.6cm}{!}{\begin{tabular}{l|l|l}
\hline 
\\*[-3.5ex] 
\multicolumn{3}{l}{\sf stellar parameter}\\[-1ex]
\hline
&&\\[-4.5mm]
{$M_\star$}   
   & stellar mass $[M_\odot]$ & 0.5, 1.0, 1.5, 2.0, 2.5\\
{$age$}
   & age [Myr] & 1, 3, 10, 100\\
{$f_{\rm UV}$}
   & excess UV $f_{\rm UV}\!=\!L_{UV}/L_\star$ & 0.001, 0.1\\
\hline
\\*[-3.5ex] 
\multicolumn{3}{l}{\sf disc parameter}\\[-1ex]
\hline
&&\\[-4mm]
{$M_{\rm d}$} 
   & disc dust mass $[M_\odot]$ & 
      \hspace*{-8mm}$10^{-7}$, $10^{-6}$, $10^{-5}$, $10^{-4}$,
      $10^{-3}$\hspace*{-2mm}\\
{$\rho_{\rm d}/\rho_{\rm g}$}\hspace*{-2mm} 
   & dust/gas mass ratio & 0.001, 0.01, 0.1, 1, 10\\
{$R_{\rm in}$} 
   & inner disc radius $[R_{\rm subli}]$ & 1, 10, 100\\
{$R_{\rm out}$} 
   & outer disc radius [AU] & 100, 300, 500\\
{$\epsilon$} 
   & column density $N_{\rm H}(r)\propto r^{\displaystyle-\epsilon}$ & 
             0.5, 1.0, 1.5 \\[-0.5ex]
{$H_0$} 
	& scale height [AU]  & {\sl fixed}: 10~@~$r_0\!=\!100$AU\\
{$\beta$} 
   & disc flaring
      $H(r)=H_0\big(\frac{r}{r_0}\big)^{\displaystyle\beta}$
      \hspace*{-3mm} & 
             0.8, 1.0, 1.2\\
\hline
\\*[-3.5ex] 
\multicolumn{3}{l}{\sf dust parameter}\\[-1ex]
\hline
&&\\[-4.5mm]
{$a_{\rm min}$} 
    & minimum grain size $[\mu m]$ & 0.05, 1\\
{$a_{\rm max}$} 
    & maximum grain size $[\mu m]$ & {\sl fixed:} 1000\\
{$s$} & settling 
    $H(r,a) \propto H(r)\;a^{\displaystyle-s/2}$
    & 0, 0.5\\
\hline
\\*[-3.5ex] 
\multicolumn{3}{l}{\sf inclination angle}\\[-1ex]
\hline
&&\\[-4mm]
$i$ & \multicolumn{2}{l}{
  $0^o$ (face-on), $41.4^o$, $60^o$, $75.5^o$, $90^o$ (edge-on)}
\end{tabular}}
\vspace*{-1mm}
\end{table}

\begin{table}
\caption{List of output quantities: monochromatic
  continuum fluxes $F_{\rm cont}$ and integrated, 
  continuum-subtracted line fluxes $F_{\rm line}$.}
\label{tab:Results}
\vspace*{-1.5mm}
\begin{tabular}{c|l}
\hline
\\*[-3.3ex] 
$\!\!\!${\sf observable}$\!\!\!$ & {\sf wavelengths} [$\mu$m]\\[-0.7ex]
\hline 
&\\[-4mm]
$F_{\rm cont}$ & 57 $\lambda$-points 
                 between $0.1\mu$m and $3500\mu$m\\
OI        & 63.18, 145.53 \\
CII       & 157.74\\
$^{12}$CO & 2600.76, 1300.40, 866.96, 650.25, 520.23, 433.56, 371.65,
          $\!\!\!$\\ 
          & 325.23, 289.12, 260.24, 144.78, 90.16,79.36, 72.84\\
o-H$_2$O  & 538.29, 179.53, 108.07, 180.49, 174.63, 78.74\\
p-H$_2$O  & 269.27, 303.46, 100.98, 138.53, 89.99, 144.52
\end{tabular}
\vspace*{-1mm}
\end{table}
 
In the DENT grid, 11 variable and 2 fixed input parameters (see
Table~\ref{tab:Parameters}) are required to specify the star\,+\,disc
systems. The input {\bf stellar parameters} are mass and age. The
corresponding effective temperatures and luminosities are from the
evolutionary models of \citet{Siess2000}. For the photospheric
spectra, we use Kurucz stellar atmosphere models of solar abundance
and matching $T_{\rm eff}$ and $\log(g)$. Because some young stars
show significant accretion and/or chromospheric activity, an extra UV
component is added to the spectrum in DENT. This extra UV component
has an important impact on the disc chemistry and temperature. It is
defined as $f_{\rm UV}\!=\!L_{\rm UV}/L_\star$ where $L_{\rm
  UV}\!=\!\int_{\rm 91\,nm}^{\rm 250\,nm} L_\lambda\,d\lambda$ is the
UV luminosity with assumed spectral shape
$L_\lambda\!\propto\!\lambda^{0.2}$.

There are {\bf other sources of energy in the discs.} The interstellar
irradiation is assumed to be isotropic and fixed throughout the grid
by \citep[Eq.~27 of][with $\chi^{\rm ISM}\!\!=\!1$]{Woitke2009a}. The
cosmic ray ionisation rate of H$_2$ is set to $\zeta_{\rm
  CR}\!=\!1.7\!\times\!10^{-17}\rm\,s^{-1}$. We further assume that
all discs are passive, i.e. there is no viscous heating. This will
have an impact on the inner structure and temperature of some
discs. However, the SED of several sources in the literature are well
fitted without viscous heating, e.g. IM Lupi \citep{Pinte2008} and
IRAS~04158+2805 \citep{Glauser2008}. 

The {\bf density structure of the disc} is parametric, with power-laws
used for the surface density, the scale height, and the dust size
distribution. Optical properties of {\sl astronomical silicate}
\citep{Draine1984} are used to calculate the dust opacities.  

The {\bf chemical abundances} are calculated
selecting 9 atomic elements, 66 molecular and 5 ice species, 950
reactions, and element abundances as outlined in \citep{Woitke2009a}.
The numerical code used here feature an improved coupling between the
UV radiative transfer and the calculation of the UV photo-rates
\citep[see][]{Kamp2009}. This version of DENT does not contain
PAHs. This will have an impact in particular on the Herbig AeBe's,
with a possible increase of the gas temperature, especially in the
upper layers of the discs. We do not expect a significant effect on
the continuum. The role of PAHs will be explored in a forthcoming
paper.

To calculate the DENT grid, two numerical codes were used in a
sequence that we now describe (see Fig.~\ref{fig:ModelPipeline}).  In
{\bf phase~1}, \MCFOST\ solves the dust radiative transfer problem to
obtain the dust temperature structure $T_{\rm d}(r,z)$ and the
internal mean intensities $J_\nu(r,z)$.  In {\bf phase~2}, these data
are transferred to \ProDiMo\ which calculates the gas temperature
structure $T_{\rm g}(r,z)$ assuming gas thermal balance, the
chemical composition $n_{\rm mol}(r,z)$ assuming kinetic chemical
equilibrium, and the level population of the gas species in the
disc. An escape probability method is used to calculate the level
populations. Phase~2 requires an additional gas parameter: the
dust-to-gas ratio, $\rho_{\rm d}/\rho_{\rm g}$. Among the results of
phase~2 are the total species masses $M_{\rm mol}$ and averaged gas
temperatures $\langle T_{\rm g}^{\rm mol} \rangle$ for $\rm
mol\!\in\!\{O,C^+,CO,H_2O\}$, defined as
\begin{eqnarray}
  \langle T_{\rm g}^{\rm mol} \rangle &\!\!=\!\!& 
              \frac{\int n_{\rm mol}(r,z)\,T_{\rm g}(r,z)\;dV}
                   {\int n_{\rm mol}(r,z)\;dV}\ ,
  \label{eq:Tmean}
\end{eqnarray}
where $n_{\rm mol}(r,z)$ is the particle density at position $(r,z)$
in the disc. In {\bf phase~3}, the level populations are transferred back to
\MCFOST\ to calculate the emission line profiles. The formal line transfer
solutions are computed in 301 velocity bins on $100\!\times\!72$
parallel rays organised in log-equidistant concentric rings in the
image plane. A Keplerian rotation velocity field is assumed for the
bulk disk kinematics, and a thermal $\!+\!$ turbulent broadening with
$v_{\rm turb}\!=\!0.15\,$km/s is added.  The calculations are
completed by {\bf phase~4} running formal solutions of the dust continuum
radiative transfer problem on the same rays. The
calculated continuum and line intensities are post-processed to get
the integrated line fluxes after continuum subtraction. Further
details are listed in Table~\ref{tab:Results}.

In total, the DENT grid comprises 323020 disc models and SED
calculations. A total number of 1610150 line flux 
calculations have been carried out for 29 spectral lines under 5
inclinations. We note that some parameter combinations may lead
to unrealistic models, but they have been kept for the sake of
completeness.

\section{Results}
\label{results}

\begin{figure}
  \centering
  \hspace*{-1mm}\includegraphics[width=8.6cm,height=6.3cm]
                  {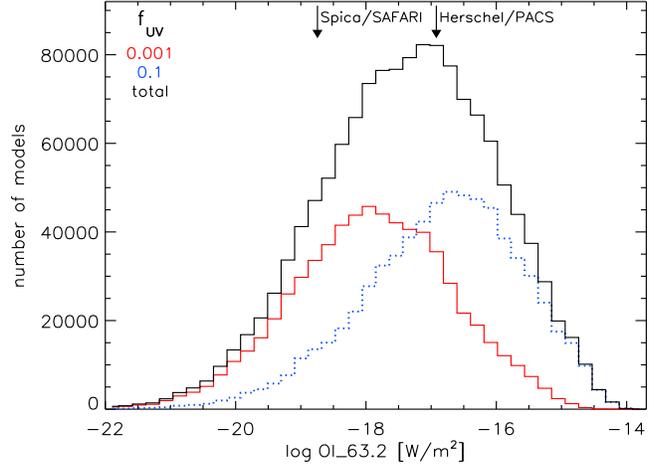}\\[-1mm]
  \caption{Dependence of line flux [OI]\,63.2$\mu$m on the stellar
  UV excess $f_{\rm UV}$. The black histogram counts the DENT models
  that result in certain [OI]\,63.2$\mu$m fluxes at distance 140pc in 40
  log-equidistant bins. The red histogram represents the low $f_{\rm
  UV}\!=\!0.001$ models, and the blue dotted histogram the high $f_{\rm
  UV}\!=\!0.1$ models. The difference between high and low UV excess
  causes a difference of about $1-1.5$ orders of magnitude in line
  flux. The arrows show the $(3\sigma, 0.5h)$ detection limits of 
  {\sc Spica/Safari} and {\sc Herschel/Pacs}, see dependence on disc mass 
  in Table~\ref{tab:detection}.}
  \label{fig:fUV}
  \vspace*{-1mm}
\end{figure}

Figure~\ref{fig:fUV} depicts all calculated line fluxes of
$\rm[OI]\,63.2\,\mu$m in form of 3 histograms, underlining the
depending on the assumed stellar UV excess. Models with high $f_{\rm
UV}$ (blue in the figure) have warm disc surfaces {heated by the
stellar UV} with $\Tg\!\gg\!\Td$, and hence strong emission
lines. Models with low $f_{\rm UV}$ (red) have gas temperatures more
equal to the dust temperatures, and hence less strong emission
lines. The dependence of $F_{\rm line}$ on $f_{\rm UV}$ is significant
for all stars (also in other lines), but is less pronounced for Herbig
Ae/Be stars which already produce photospheric soft UV, even for $f_{\rm
UV}\!=\!0$. For example, a Kurucz stellar atmosphere model with
$T_{\rm eff}\!=\!8500\,$K, $\log(g)\!=4$ attains $L_{\rm
UV}\!\approx\!0.097L_\star$ whereas a model with $T_{\rm eff}\!=\!4500\,$K,
$\log(g)\!=4$ only attains $L_{\rm
UV}\!\approx\!2\cdot10^{-6}L_\star$. Table~\ref{tab:detection} lists
the fraction of models predicting line fluxes of [OI] and [CII] larger
than (0.5h\,,\,3$\sigma$) detection limits of {\sc Herschel/Pacs} and
{\sc Spica/Safari} at 140pc as function of disc gas mass. {Note that,
for massive discs, the line emitting regions may be optically thick in
the continuum and may have $T_{\rm g}\!\approx\!T_{\rm d}$. In
such cases, there is only very limited contrast between line and
continuum, leaving a fair fraction of massive discs with
non-detectable lines.}

\newcommand{\z}[0]{\hspace*{-0.8mm}}
\begin{table}
\vspace*{-1mm}
\caption{Fraction of DENT models predicting line fluxes larger than
  (0.5h\,,\,3$\sigma$) detection limits of {\sc Herschel/Pacs} and
  {\sc Spica/Safari} at 140pc as function of total disc gas mass
  $M_{\rm gas}\rm\,[M_\odot]$ and stellar UV excess $f_{\rm UV}$. Each  entry in this table is based on more than 5000 disc models. We
  neglect detection problems due to background confusion here.}
\label{tab:detection}
  \vspace*{-1.5mm}\hspace*{-1mm}\resizebox{8.1cm}{!}{
  \begin{tabular}{c|ccccccc}
   \hline
   \\*[-3.3ex] 
  \hspace*{-2mm}$_{f_{\rm UV}}$ 
  \hspace*{-2mm}\resizebox{3.5mm}{2.5mm}{\textbackslash}\hspace*{-2mm}
  $^{M_{\rm gas}}$\hspace*{-2mm}
               & \z$10^{-7}$\z & \z$10^{-6}$\z & \z$10^{-5}$\z &
                 \z$10^{-4}$\z & \z$10^{-3}$\z & \z$10^{-2}$\z & 
                 \z$10^{-1}$\z \\[-0.3mm]
   \hline
   \hline
   \\*[-3.3ex] 
   \multicolumn{8}{l}{$|F_{\rm [OI]63.2\mu{\rm m}}|
             >1.2\!\times\!10^{-17}\rm\,W/m^2$ (\sc Herschel/Pacs)}\\*[0.5mm] 
   0.1\z       & \z0\%\z  & \z20\%\z & \z51\%\z & \z66\%\z & 
                 \z70\%\z & \z71\%\z & \z71\%\z \\
   0.001\z     & \z0\%\z  & \z6\%\z  & \z17\%\z & \z23\%\z & 
                 \z26\%\z & \z30\%\z & \z34\%\z \\*[-0.5mm] 
   \hline
   \\*[-3.3ex] 
   \multicolumn{8}{l}{$|F_{\rm [OI]145.5\mu{\rm m}}|
               >4\!\times\!10^{-18}\rm\,W/m^2$ (\sc Herschel/Pacs)}\\*[0.5mm] 
   0.1\z       & \z0\%\z  & \z0\%\z  & \z9\%\z  & \z30\%\z & 
                 \z45\%\z & \z52\%\z & \z57\%\z \\
   0.001\z     & \z0\%\z  & \z0\%\z  & \z3\%\z  & \z7\%\z & 
                 \z12\%\z & \z18\%\z & \z23\%\z \\*[-0.5mm] 
   \hline
   \\*[-3.3ex] 
   \multicolumn{8}{l}{$|F_{\rm [CII]157.7\mu{\rm m}}|
               >4\!\times\!10^{-18}\rm\,W/m^2$ (\sc Herschel/Pacs)}\\*[0.5mm] 
   0.1\z       & \z0\%\z  & \z0\%\z  & \z17\%\z & \z52\%\z & 
                 \z56\%\z & \z57\%\z & \z56\%\z \\
   0.001\z     & \z0\%\z  & \z0\%\z  & \z6\%\z  & \z14\%\z & 
                 \z14\%\z & \z14\%\z & \z13\%\z \\*[-0.5mm] 
   \hline
   \hline
   \\*[-3.3ex]    
   \multicolumn{8}{l}{$|F_{\rm [OI]63.2\mu{\rm m}}|
               >1.8\!\times\!10^{-19}\rm\,W/m^2$ (\sc Spica/Safari)}\\*[0.5mm] 
   0.1\z       & \z65\%\z & \z90\%\z & \z96\%\z & \z95\%\z & 
                 \z96\%\z & \z97\%\z & \z96\%\z \\
   0.001\z     & \z38\%\z & \z65\%\z & \z79\%\z & \z83\%\z & 
                 \z86\%\z & \z86\%\z & \z85\%\z \\*[-0.5mm] 
   \hline
   \\*[-3.3ex]    
   \multicolumn{8}{l}{$|F_{\rm [OI]145.5\mu{\rm m}}|
               >1.2\!\times\!10^{-19}\rm\,W/m^2$ (\sc Spica/Safari)}\\*[0.5mm] 
   0.1\z       & \z5\%\z  & \z44\%\z & \z68\%\z & \z80\%\z & 
                 \z85\%\z & \z87\%\z & \z88\%\z \\
   0.001\z     & \z1\%\z  & \z15\%\z & \z29\%\z & \z46\%\z & 
                 \z56\%\z & \z62\%\z & \z63\%\z \\*[-0.5mm] 
   \hline
   \\*[-3.3ex]    
   \multicolumn{8}{l}{$|F_{\rm [CII]157.7\mu{\rm m}}|
               >1.2\!\times\!10^{-19}\rm\,W/m^2$ (\sc Spica/Safari)}\\*[0.5mm] 
   0.1\z       & \z0\%\z  & \z83\%\z & \z95\%\z & \z94\%\z & 
                 \z95\%\z & \z94\%\z & \z93\%\z \\
   0.001\z     & \z0\%\z  & \z53\%\z & \z81\%\z & \z78\%\z & 
                 \z76\%\z & \z73\%\z & \z69\%\z \\ 
  \end{tabular}}
\vspace*{-1mm}
\end{table}

\begin{figure}
  \centering
  \hspace*{-2.5mm}\includegraphics[width=8.8cm,height=7.2cm]
                  {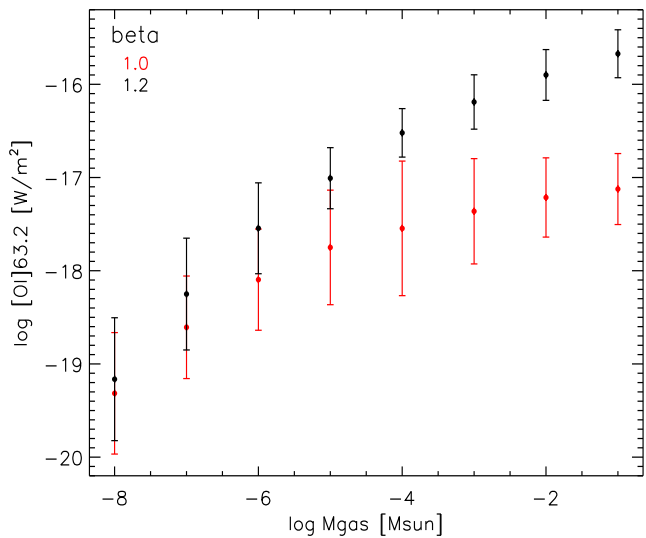}\\[-1mm]
  \caption{Dependence of line flux of [OI] 63.2$\mu$m at distance
    $d\!=\!140\,$pc on the flaring parameter $\beta$, as function of
    total disc gas mass $M_{\rm gas}$. A sub-selection of 3456
    T\,Tauri models is plotted ($M_\star\!\leq\!1\rm\,M_\odot$,
    $age\!\leq\!1\rm\,Myr$, $f_{\rm UV}\!=\!0.1$, $R_{\rm
    in}\!=\!R_{\rm subli}$, $s\!=\!0$, inclination angle $\leq\!60^o$,
    where the statistical range of line flux predictions due to the
    variation of the other input parameters in expressed by mean
    values and standard deviations.}
  \label{fig:beta}
  \vspace*{-1mm}
\end{figure}

Another clear trend in the models is depicted in Fig.~\ref{fig:beta},
which shows the calculated line fluxes of $\rm[OI]\,63.2\,\mu$m for a
sub-selection of T\,Tauri disc models with high $f_{\rm UV}$ as
function of disc mass. At low disc masses, the discs are optically
thin and the flaring of the disc (see definition of $\beta$ in
Table~\ref{tab:Parameters}) has only little influence on the line
fluxes.  However, with increasing disc mass, the inner disc becomes
optically thick and the computed line fluxes split up into two
branches. For strong flaring ($\beta\!=\!1.2$), the fluxes of the
emission lines steadily increase further, whereas for non-flaring
discs ($\beta\!\leq\!1$) they saturate at $M_{\rm
  gas}\!\approx\!10^{-5}-10^{-4}\rm\,M_\odot$, henceforth called the
``saturation disc mass''. In other words, for massive T\,Tauri discs,
high far IR line fluxes (e.g.\ $\rm[OI]\,63.2\,\mu m\!\ga\!3\times
10^{-17}\,W/m^2$) are a safe indicator of disc (gas) flaring.
In flared geometry, the disc surface is directly heated by the star,
hence higher temperatures and stronger emission lines. In non-flared
geometry, the surface layers are situated in the shadow casted by the
dust in the inner disc regions, and the gas temperatures are
cooler. In that case, a further increase in disc mass does not lead to
stronger emission lines, but rather to an increase of the shielding
effects, causing cooler temperatures and sometimes even weaker
emission lines.

The saturation behaviour of the emission lines for non-flared geometry
depends on the line properties. High-excitation lines like
$\rm[OI]\,145.5\,\mu$m react more sensitively to temperature changes and
hence to disc flaring, whereas low excitation lines like
CO~$J$=1$\to$0, $\rm [CII]\,157.7\,\mu$m are less affected. However,
the point where these saturation effects start to appear, the
saturation disc mass, is found to be rather similar for all lines,
because it is the amount of dust and its opacity in the inner disc
regions that causes the shielding.

\begin{figure}
  \centering
  \hspace*{-5mm}\includegraphics[width=9.05cm,height=8.0cm]
                  {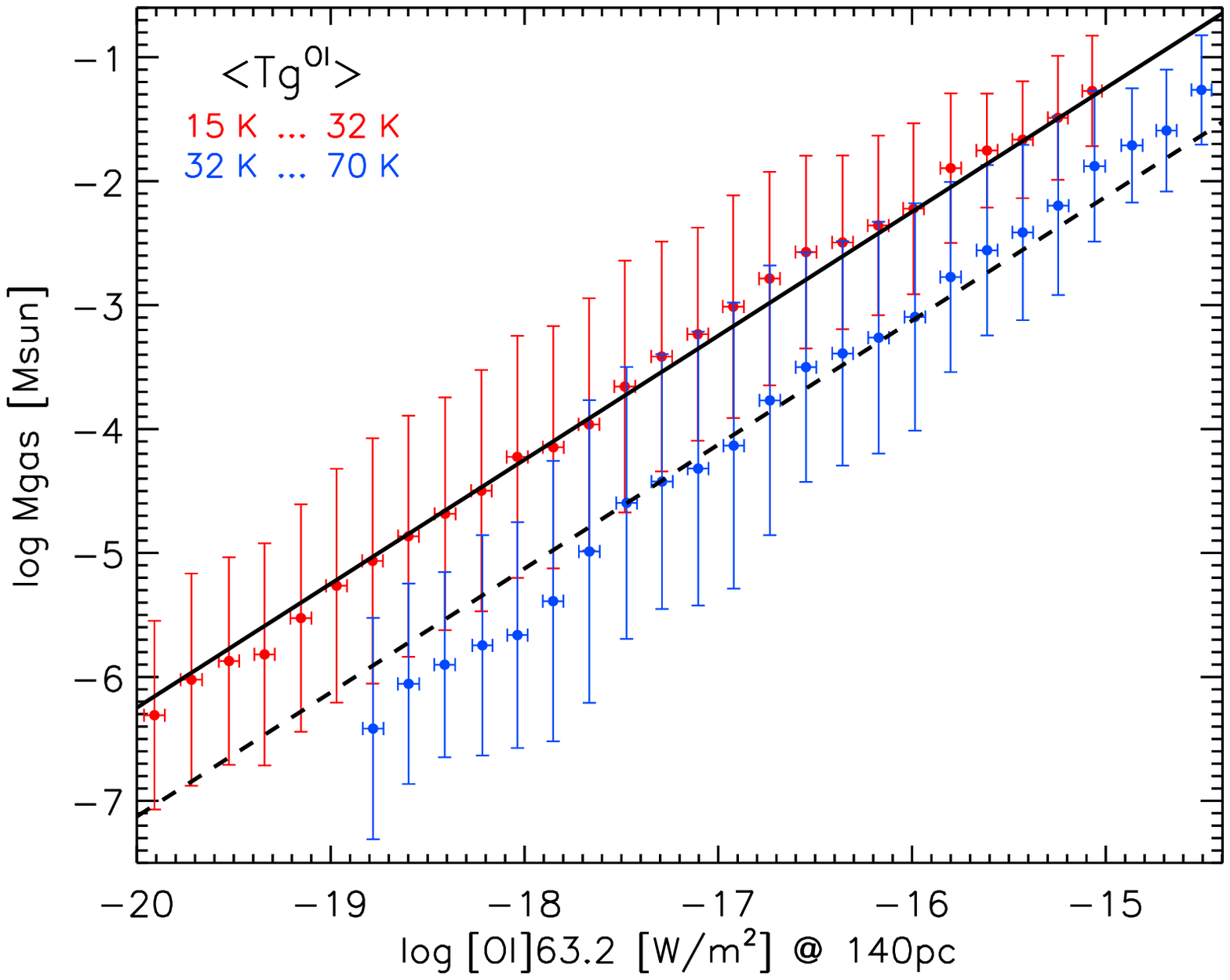}\\[-36.5mm]
  \hspace*{47mm}\includegraphics[width=34mm]
                  {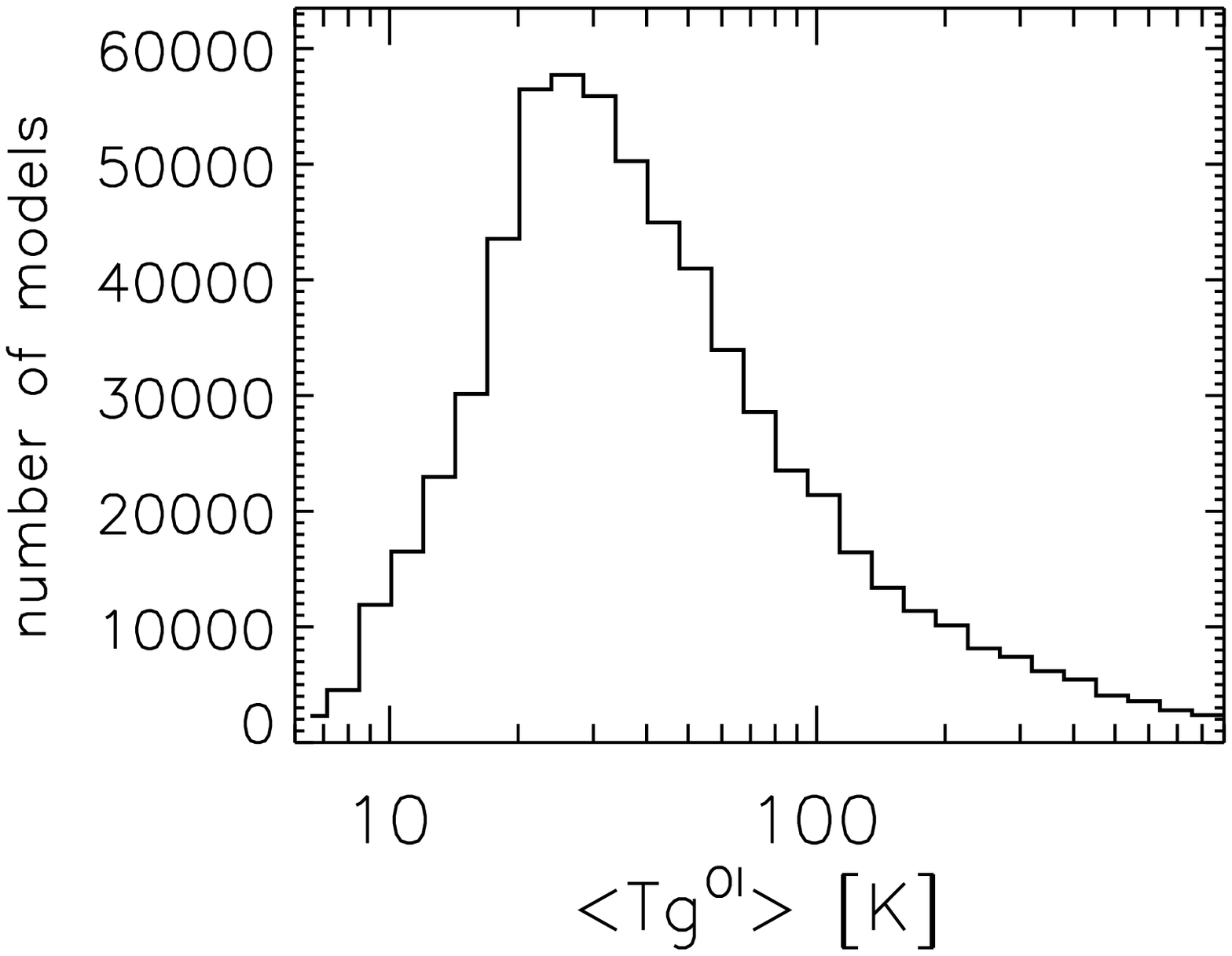}\\[9mm]
  \caption{Relation between total disc gas mass and
  $\rm[OI]\,63.2\mu$m line flux at distance 140pc. The $x$-axis is
  subdivided into 30 log-equidistant bins. In each bin, the values of
  the disc gas mass (parameter, see Table~\ref{tab:Parameters}) of all
  models with matching $\rm[OI]\,63.2\mu$m line flux are statistically
  analysed to yield a mean value and a standard variation, which are
  then plotted as points with vertical error bars. We distinguish
  between cool and warm models by $\langle T_{\rm g}^{\rm OI}\rangle$
  as defined in Eq.\,(\ref{eq:Tmean}), and have chosen temperature
  interval boundaries to bracket the $1\sigma$ distribution as shown
  by the inserted histogram. The following models have been selected:
  dust/gas ratio $\rho_{\rm d}/\rho_{\rm g}\!=\!0.01-1$, flaring index
  $\beta\!\geq\!1$, and inclination angle $\leq\!60^o$, altogether
  122269 cool and 101244 warm models. The full and dashed lines show
  the results from the formula Eq.\,(\ref{eq:Fsimple}) with $T_{\rm
  exc}\!=\!24$K and $T_{\rm exc}\!=\!30.5$K, respectively.}
  \label{fig:Mgas}
  \vspace*{-1mm}
\end{figure}

The fact that different emission lines originate in different disc
regions, and the strong dependencies of the line fluxes on
UV excess $f_{\rm UV}$ and flaring index $\beta$ suggest that a
simple, PDR-like analysis of emission line ratios to determine the total 
disc mass is difficult. However, Fig.~\ref{fig:Mgas} shows that
most of these complicated parameter dependencies affect the line
fluxes in an indirect way, namely by changing the mean temperature of
the disc. If we plot the dependency between disc gas mass and $\rm
[OI]63.2\mu$m line flux for models with similar mean disc temperatures,
we roughly retrieve a linear relation as expected from a simple
analysis (see Appendix~\ref{appendix-1}).

The $1\sigma$ errorbars in Fig.~\ref{fig:Mgas} show, however,
that there is still a considerable variation of the $\rm[OI]63.2\mu$m 
flux among the models, even if they have similar mean disc
temperatures. Note also, that the fitting value of $T_{\rm
exc}\!=\!30.5$K is not consistent with the actual disc temperatures as
measured from the models. This may be due to the way we have defined
the mean disc temperature (Eq.\,\ref{eq:Tmean}) and/or due to included
non-LTE effects. \citet{Kamp2009} have shown that, in general, $T_{\rm
exc}<T_{\rm g}$ in the outer and upper disc layers. 

We note that an unmindful usage of Eq.\,(\ref{eq:Fsimple}) with an
assumed value for $T_{\rm exc}$ (e.g. 27K) for the purpose of gas mass
determination from a measured $\rm[OI]63.2\mu$m line flux can be
misleading and does not account for the variety of results that we
find in the DENT model grid. In particular, high mass discs tend to be
cooler as demonstrated by the saturation behaviour depicted in
Fig.~\ref{fig:beta}, and their oxygen fine-structure lines can easily
become optically thick.


\section{Summary and Conclusions}
\label{summary}

In a concerted effort of the theory groups in Edinburgh, Grenoble and
Groningen, we have computed a grid of 300\,000 circumstellar disc
models, {simultaneously solving} gas-phase, UV-photo and ice chemistry, 
detailed heating \& cooling balance, and continuum \& line radiative
transfer.

The first results of the DENT grid show a strong dependence of the
calculated emission line fluxes on the assumed stellar UV excess
and on the flaring of the disc. The stellar UV is
essential for the heating of the upper disc layers. In combination
with positive disc flaring, a strong stellar UV irradiation creates an
extended warm surface layer with $\Tg\!>\!\Td$ responsible for the
line emissions. However, if the disc is not flared (self-shadowed),
discs with total mass $\ga\!10^{-5}-10^{-4}\rm\,M_{\odot}$ increasingly
shield the stellar UV by their inner parts, which causes much cooler
surface layers, and a saturation of the line fluxes with increasing
disc mass.

Despite these complicated parameter dependencies, we have shown that
the $\rm[OI]\,63.2\mu$m line flux depends basically on two quantities,
namely the total disc gas mass and the mean disc temperature. We will
continue this work by two follow-up papers (Kamp et al.~2010, in
prep., M{\'e}nard et al.~2010, in prep.) that will provide more
insight into the statistical behaviour of gas line and dust continuum 
predictions, respectively, to identify trends and robust correlations with
disc mass.

\smallskip\noindent In summary, the DENT grid allows to\\[-0.5ex]

\hspace*{-11mm}\begin{tabular}{cp{80mm}}
$\bullet$\hspace*{-2.5mm} & study the effects of stellar, disc, and dust 
            parameters on continuum and line observations,\\
$\bullet$\hspace*{-2.5mm} & allow for a qualified interpretation of
            observational data,\\
$\bullet$\hspace*{-2.5mm} & quickly predict line and continuum fluxes
            for planning observations,\\
$\bullet$\hspace*{-2.5mm} & search for best-fitting models concerning a 
            given set of observed line and continuum fluxes,\\
$\bullet$\hspace*{-2.5mm} & study the robustness of certain fit values
            against variation of the observational data.\\[1.5ex]
\end{tabular}

\noindent We intend to make the calculated DENT grid available to the
scientific community. A graphical user interface called {\tt xDENT}
has been developed to allow researchers to visualise the DENT results,
to make plots as presented in this letter, and to search for
best-fitting models for a given set of continuum and line flux
data. We emphasise, however, that the DENT grid has not been developed for 
detailed fitting of individual objects. The coarse 
sampling of the 12-dimensional parameter space can mostly be used to 
narrow down the parameter range for individual objects, for example to 
design a finer sub-grid, especially for not so well-known objects.

With a comprehensive data set of far IR gas emission lines to be
obtained by Herschel/GASPS very soon, we aim at breaking the
degeneracy of SED fitting and make possible a more profound analysis
of the physical, chemical and temperature structures of discs around
young stars.

\appendix
\section{Simple line emission model}
\label{appendix-1}

Let us assume that an emission line is optically thin and that the
emitting species is populated with a uniform excitation temperature
$T_{\rm exc}$. The line luminosity is then given by
\begin{equation}
  L_{\rm line} = h\nu\,A_{ul}\,N_{\rm tot} \frac{g_u \exp(-E_u/kT_{\rm exc})}
                                                {Q(T_{\rm exc})} \ ,
\end{equation}
where $\nu$ is the line centre frequency, $A_{ul}$ is the Einstein
coefficient of the line transition from level $u$ to level $l$, $Q$ is
the partition function, and $g_u$ and $E_u$ are the statistical weight
and energy [K] of the upper level. The total number of line emitting
particles $N_{\rm tot}$ is related to the total disc gas mass by
\begin{equation}
  N_{\rm tot} = \epsilon N_{\langle H\rangle}
              = \epsilon\,\frac{M_{\rm gas}}{\mu_H} \ ,
\end{equation}
where $\epsilon$ is the abundance of the line emitting species with
respect to hydrogen nuclei and $\mu_H\!\approx\!1.31\,$amu the gas
mass per hydrogen nucleus, assuming solar abundances.

The observable line flux $\rm[erg/s/cm^2]$ at distance $d$ is
\begin{equation}
  F_{\rm line} = \frac{L_{\rm line}}{4\pi\,d^2} \ .
  \label{eq:Fsimple}
\end{equation}
For the case of the $\rm[OI]\,63.2\mu$m fine-structure line, we have
$u\!=\!2$, $l\!=\!1$, $A_{21}\!=\!8.87\times 10^{-5}\rm\,s^{-1}$,
$E_2\!=\!227.7\,$K, $g_1\!=\!5$ and $g_2\!=\!3$. We calculate the
partition function including the third level $E_3\!=\!326.0\,$K,
$g_3\!=\!1$. The oxygen abundance assumed in the models is
$\epsilon\!=\!8.5\times 10^{-4}$. However, the actual abundance of the
neutral oxygen atom is reduced by CO, CO-ice, H$_2$O and H$_2$O-ice
formation, and we use a mean value from the models,
$\epsilon\!=\!3.6\times 10^{-4}$.

\section*{Acknowledgements}
{\footnotesize We acknowledge financial support by ANR of France
  through contract ANR-07-BLAN-0221 (DustyDisks). We also thank
  Programme PNPS of CNRS/INSU France for supporting this work since
  the beginning. The calculations presented in this paper were made on
  the FOSTINO computer cluster, acquired as part of ANR project
  DustyDisks and operated by Service Commun de Calcul Intensif (SCCI)
  of Observatoire de Grenoble (OSUG).  WFT is supported by a Scottish
  Universities Physics Alliance (SUPA) fellowship in astrobiology.
  C. Pinte acknowledges the funding from the European Commission's
  Seventh Framework Program as a Marie Curie Intra-European Fellow
  (PIEF-GA-2008-220891).}

\bibliographystyle{mn2e}
\bibliography{reference}  
\label{lastpage}

\end{document}